\documentclass[aps,prl,twocolumn,superscriptaddress,floatfix,nofootinbib,showpacs]{revtex4}

\usepackage{color}
\usepackage[dvips]{graphicx}

\newcommand{\lsim}{\mathrel{\mathop{\kern 0pt \rlap
  {\raise.2ex\hbox{$<$}}}
  \lower.9ex\hbox{\kern-.190em $\sim$}}}
\newcommand{\gsim}{\mathrel{\mathop{\kern 0pt \rlap
  {\raise.2ex\hbox{$>$}}}
  \lower.9ex\hbox{\kern-.190em $\sim$}}}

\newcommand{\be}{\begin{equation}}
\newcommand{\ee}{\end{equation}}
\newcommand{\beqarr}{\begin{eqnarray}}
\newcommand{\eeqarr}{\end{eqnarray}}

\newcommand{\vmin}{v_{\rm min}}
\newcommand{\vesc}{v_{\rm esc}}
\newcommand{\ivmin}{{\cal I}(\vmin)}

\begin{document}

\title{Temporal distortion of annual modulation at low recoil energies}


  
%
\author{N. Fornengo}
\affiliation{Dipartimento di Fisica Teorica, Universit\`a di Torino \\
Istituto Nazionale di Fisica Nucleare, Sezione di Torino \\
via P. Giuria 1, I--10125 Torino, Italy}

\author{S. Scopel} 
\affiliation{Dipartimento di Fisica Teorica, Universit\`a di Torino \\
Istituto Nazionale di Fisica Nucleare, Sezione di Torino \\
via P. Giuria 1, I--10125 Torino, Italy}

\date{\today}

\begin{abstract}
We show that the main features of the annual modulation of the signal
expected in a WIMP direct detection experiment, i.e. its sinusoidal
dependence with time, the occurrence of its maxima and minima during
the year and (under some circumstances) even the one--year period, may
be affected by relaxing the isothermal sphere hypothesis in the
description of the WIMP velocity phase space. The most relevant effect
is a distortion of the time--behaviour at low recoil energies for
anisotropic galactic halos. While some of these effects turn out to be
relevant at recoil energies below the current detector thresholds,
some others could already be measurable, although some degree of
tuning between the WIMP mass and the experimental parameters would be
required. Either the observation or non--observation of these effects
could provide clues on the phase space distribution of our galactic
halo.
\end{abstract}


\maketitle

\section{Introduction}
\label{sec:intro}

It is widely believed, as suggested by a host of independent
cosmological and astrophysical observations, that the most part of the
matter in the Universe is not visible, revealing its existence only
through gravitational effects. In particular, data on the rotational
curves of galaxies indicate that the galactic visible parts are
surrounded by approximately--spherical dark halos which extend up to
several times the size of the luminous components.

The best candidates to provide dark matter in galaxies are Weakly
Interacting Massive Particles (WIMP). Several WIMP direct--detection
experiments are operating \cite{direct}, with the goal of measuring
the nuclear recoil energy (in the KeV range) expected to be deposited
in solid, liquid or gaseous targets by the scattering of the
non-relativistic dark halo WIMPs. Unfortunately, expected rates are
small and the exponential decay of the WIMP recoil--spectrum resembles
that of the background at low energies. However, a specific signature
can be exploited in order to disentangle a WIMP signal from the
background: the annual modulation of the rate \cite{modulation}. This
effect, expected to be of the order of a few per cent, is induced by
the rotation of the Earth around the Sun. Due to its smallness, the
annual modulation signature requires large--mass detectors with high
statistics in order to overcome background fluctuations and be
unambiguously detected. The annual modulation effect has been
experimentally investigated by the DAMA Collaboration, which has
indeed reported a positive evidence by using a 100 kg sodium iodide
detector \cite{damalast}.

One of the most important sources of uncertainty in the calculation of
WIMP direct detection rates is the modeling of the velocity
distribution function (DF) of the particles populating the dark
halo. In the literature, a simple isothermal sphere model is usually
adopted, i.e. a WIMP gas described by an isotropic Maxwellian with
r.m.s velocity of the order of 300 Km s$^{-1}$. This leads to a
sinusoidal time--dependence of the expected signal with maximum (or
minimum) around June 2$^{\rm nd}$, i.e. with the same (or opposite)
phase as the relative velocity between the Earth and the halo rest
frame.

However, the actual form of the WIMP velocity DF is unknown, and many
different models, alternative to the isothermal sphere, are compatible
with observations \cite{galaxy}. The goal of the present Letter is to
show that the main features of the annual modulation effect (the
sinusoidal dependence with time, the occurrence of maxima and minima
during the year and, under some circumstances, the one--year period)
may be affected by anisotropies in the velocity DF. The most relevant
effect is a distortion of the sinusoidal time--behaviour at low recoil
energies. These energies, though below the current detector
thresholds, might be reached in the future. The observation of the
effects discussed in this Letter could provide informations on the
phase space distribution of our galactic halo, especially on the
degree of its anisotropy.

\section{The annual modulation effect}  

Due to the rotation of the disk around the galactic center, the
solar system moves through the WIMP halo, assumed to be at rest in the
galactic rest frame. In the following, we will assume a right--handed
system of orthogonal coordinates: the $x$ axis in the galactic plane,
pointing radially outward; the $y$ axis in the galactic plane,
pointing in the direction of the disk rotation; the $z$ axis directed
upward, perpendicular to the galactic plane. Notice that our system
differs from standard ``galactic coordinates'' by the different choice
of the $x$ axis, which for us is directed outward.

The relative velocity between the WIMP halo and the detector is given
by the Earth velocity $\vec v^E$, as seen in the galactic rest
frame. It is the sum of three components: the galactic rotational
velocity $\vec v^G = (0,v_0,0)$ Km s$^{-1}$ (we will assume: $v_0 =
220$ Km s$^{-1}$), the Sun proper motion $\vec v^S = (-9,12,7)$ Km
s$^{-1}$\cite{langlsmith} and the Earth orbital motion $\vec
u^E(t)$\cite{langlsmith}:
\begin{eqnarray}
v^E_x &=& v^G_x + v^S_x + u^E(\lambda) \cos\beta_x \cos[\omega (t-t_x)] \label{eq:vearthx}, \\
v^E_y &=& v^G_y + v^S_y + u^E(\lambda) \cos\beta_y \cos[\omega (t-t_y)] \label{eq:vearthy}, \\
v^E_z &=& v^G_z + v^S_z + u^E(\lambda) \cos\beta_z \cos[\omega (t-t_z)] \label{eq:vearthz},
\end{eqnarray}
where $\lambda$ is the ecliptic longitude, which is function of time.
We can express $\lambda$ as \cite{langlsmith}: $\lambda = L+
{1^\circ}.915 \sin g + {0^\circ}.020 \sin 2g$ where
$L={280^\circ}.460+{0^\circ}.9856474~t$ and $g =
{357^\circ}.528+{0^\circ}.9856003~t$ and $t$ denotes the time
expressed in days relative to UT noon on December 31. In
Eqs.(\ref{eq:vearthx}--\ref{eq:vearthz}) $u^E(\lambda) =
\langle{u^E}\rangle [1-e \sin(\lambda - \lambda_0)]$ is the modulus of
the Earth rotational velocity, which slightly changes with time due to
the small ellipticity $e$ of the Earth orbit
($\langle{u^E}\rangle=29.79$ Km s$^{-1}$, $e=0.016722$ and
$\lambda_0=13^\circ \pm 1^\circ$) \cite{langlsmith}. In
Eqs.(\ref{eq:vearthx}--\ref{eq:vearthz}) the $\beta_i$ denote the
ecliptic latitudes and the $t_i$ are the phases of the three velocity
components: $\beta_x={174^\circ}.4697$,
$\beta_y={59^\circ}.575$, $\beta_z={29^\circ}.812$ and $t_x=76.1$ day,
$t_y=156.3$ day, $t_z=352.4$ day. The angular velocity has a period of
1 year, and is given by: $\omega = 2\pi/(365~{\rm days})$.  With the
numbers given above, the modulus of the Earth velocity changes in time
as: $v_E\equiv |\vec v_E| \simeq 233.5 + 14.4 \cos[\omega(t-t_0)]$ (in
Km s$^{-1}$), where $t_0\simeq 152$ days, i.e. June 2$^{\rm
nd}$. Notice the slight offset between $t_0$ and $t_y$, due to the
composition of the velocity components. As a good approximation, $v_E$
is usually taken as: $v_E = (v_0 + v^S_y) + \langle u^E \rangle
\cos\beta_y \cos[\omega(t-t_0)]$. 

The direct detection differential rate $dR/dE_R$ is proportional to
the integral:
\begin{equation}
\ivmin = \int_{w \geq \vmin} d\vec w \;\;\frac{f_{\rm ES}(\vec w)}{w},
\label{eq:ivmin}
\end{equation}
where $f_{\rm ES}$ and $w$ are the WIMP velocity DF and the WIMP
velocity, respectively, in the Earth's rest frame; $\vmin$ is the
minimum value of $w$ for a given recoil energy $E_R$, WIMP mass $m_W$
and nuclear target mass $m_N$ and is given by: $\vmin \equiv
\sqrt{E_R/(2 m_N)}(m_W+m_N)/m_W$.

By indicating with $f(\vec v)$ and $\vec v$ the WIMP velocity DF and
the WIMP velocity in the galactic reference frame, the following
transformations hold:
\begin{eqnarray}
\vec v &\rightarrow & \vec w = \vec v - \vec v^E(t),\\
f(\vec v) &\rightarrow & f_{\rm ES}(\vec w)=f(\vec w + \vec v^E(t)),
\label{eq:transformation}
\end{eqnarray}
which imply that $\ivmin$, and so $dR/dE_R$, develops a time dependence
induced by $\vec v_E (t)$.

\section{The isotropic isothermal sphere}

\begin{figure}[t] \centering
\vspace{-30pt}\includegraphics[width=1.0\columnwidth]{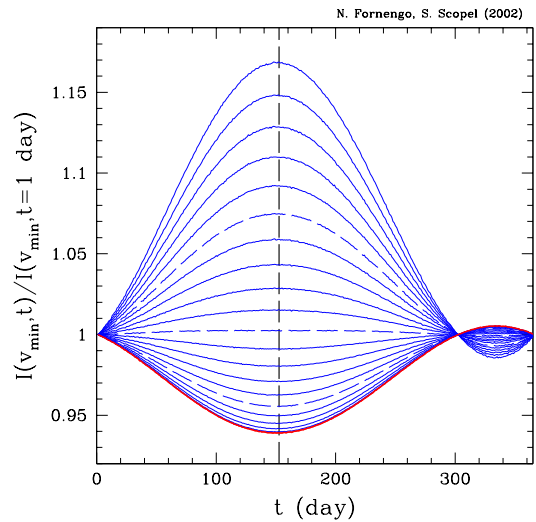}
\vspace{-30pt}\caption{\label{fig:ivmin_isotropic} Time dependence of $\ivmin$ for an
isotropic isothermal sphere. The different curves refer to values of
$\vmin$ ranging from 0 (lower) to 400 (upper) Km s$^{-1}$, in steps of
20 Km s$^{-1}$ (the dashed curves correspond to $\vmin=100$, 200 and
300 Km s$^{-1}$). The vertical dashed line denotes $t=152$ days. }
\end{figure}

In the case of the isothermal sphere model the DF is given by a
truncated isotropic Maxwellian which depends only on $v\equiv|\vec
v|$:
\begin{equation}
f(v) = N \exp\left(-\frac{v^2}{2 \sigma^2}\right)
\left |\phantom{\frac{1}{2}}\right._{v\leq \vesc},
\end{equation}
where $\sigma$ is the WIMP r.m.s. velocity, given by:
$\sigma^2=3 v_0^2/2$. It is clear that, through the change of
reference frame of Eq. (\ref{eq:transformation}), $\ivmin$ depends on
time only through $v_E$. Since the relative change of $v_E$ during the
year is of the order of a few per cent, we can approximate $\ivmin$
with its first--order expansion in the small parameter $\epsilon\equiv
\delta v_E/v_E$, around its mean value ${\cal I}_0 (\vmin)$: $\ivmin -
{\cal I}_0(\vmin) \propto \cos[\omega (t-t_0)]$.  The well known
result is then obtained that the WIMP rate has a sinusoidal time
dependence with the same phase ($t_0\simeq$ June 2$^{\rm nd}$) as
$v_E$, for all values of $\vmin$. This is shown in
Fig.\ref{fig:ivmin_isotropic}, where $\ivmin$ is plotted as a function
of time for various values of $\vmin$.

\section{Anisotropic models}

The simplest generalization of the isothermal sphere model is given by
a triaxial system described by a multivariate gaussian:
\begin{equation}
f(\vec v) = N \exp\left(-\frac{v_x^2}{2 \sigma_x^2}-\frac{v_y^2}{2
\sigma_y^2}-\frac{v_z^2}{2 \sigma_z^2}\right) 
\left |\phantom{\frac{1}{2}}\right._{v\leq \vesc}, 
\label{eq:triaxialGS}
\end{equation}
where $N$ is the normalization constant. For $\vesc \rightarrow
\infty$, then $N=[(2\pi \sigma_x^2)(2\pi \sigma_y^2)(2\pi
\sigma_z^2)]^{-1/2}$.  The usual isothermal sphere is the spherical
limit of Eq.(\ref{eq:triaxialGS}), obtained with:
$\sigma_x=\sigma_y=\sigma_z \equiv \sigma$. In order to discuss the
effect of anisotropy at fixed WIMP mean kinetic energy, we will fix
$\sigma^2=\sigma_x^2+\sigma_y^2+\sigma_z^2$ as in the isothermal case
($\sigma^2=3 v_0^2/2$) and discuss our results in terms of the two
independent parameters: $\lambda_{12}\equiv\sigma_x/\sigma_y$ and
$\lambda_{32}\equiv\sigma_z/\sigma_y$.

At variance with the isothermal sphere, now $\ivmin$ depends in
general on all the three components of $\vec v_E$, and not simply on
$v_E$. We can write:
\begin{equation}
f_{\rm ES}(\vec w) = N \exp\left[-(\vec r + \vec \eta)^2\right],
\label{eq:triaxialES}
\end{equation}
where we have defined the reduced (dimensionless) variables: $r_i =
w_i/(\sqrt{2}\sigma_i)$ and $\eta_i = v^E_i/(\sqrt{2}\sigma_i) =
\eta^0_i + \delta \eta_i \cos[\omega (t-t_i)]$ ($i=x,y,z$). In the
isotropic case (isothermal sphere) one has $\eta^0_{x,z}\ll 1$,
$\eta^0_y \sim 1$ and $\delta\eta_i \ll 1$. The presence of the
order--one parameter $\eta^0_y$ and of the small oscillation
amplitudes $\delta\eta_i$ allows the Taylor--expansion of $\ivmin$ in
terms of $\epsilon_i\equiv\delta\eta_i/\eta^0_y$
parameters. A straightforward calculation shows that the conclusions
of the previous Section are recovered.

\begin{figure}[t] \centering
\vspace{-30pt}\includegraphics[width=1.0\columnwidth]{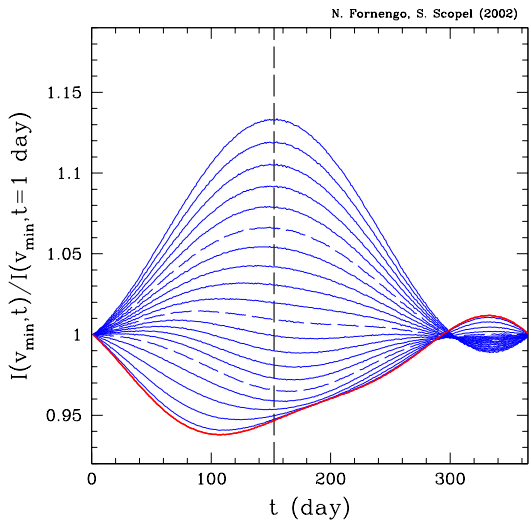}
\vspace{-30pt}\caption{\label{fig:ivmin_tangential} The same as in
Fig. \ref{fig:ivmin_isotropic}, for an anisotropic model with
$\lambda_{12}\equiv\sigma_x/\sigma_y = 0.2$ and
$\lambda_{32}\equiv\sigma_z/\sigma_y = 0.8$.}
\end{figure}

On the contrary, allowing anisotropies such that $\lambda_{12}<1$
and/or $\lambda_{32}<1$, the values of the parameters $\eta_{x,z}$ and
$\delta\eta_{x,z}$ are enhanced, and the time--dependence of $v^E_x$
and $v^E_z$ in $\ivmin$ may become important. An example of this
situation is shown in Fig. \ref{fig:ivmin_tangential}, where, the time
evolution of $\ivmin$ is plotted for $\lambda_{12}=0.2$ and
$\lambda_{12}=0.8$ (i.e.: $(\sigma_x,\sigma_y,\sigma_z) =
(42,208,166)$ Km s$^{-1}$.)  This choice of $\lambda_{12}$,
$\lambda_{32}$, which refers to a tangential anisotropy, corresponds
to triaxial models discussed, for instance, in Refs.
\cite{carollo,galaxy}. A distortion of the curves of
Fig. \ref{fig:ivmin_tangential}, as compared to the familiar
sinusoidal time--dependence, appears: this effect may be explained by
the fact that now $\delta\eta_x \sim 1$, a Taylor--expansion of the
type used in the isothermal sphere case breaks down and a full
numerical calculation of the integral of Eq.(\ref{eq:ivmin}) is
required. The final result is not sinusoidal. This peculiar behaviour
is more pronounced at low values of $\vmin$ (i.e. low recoil energies)
namely for $\vmin\lsim$ 80 Km s$^{-1}$, for which the distortion is
strong and the maxima (in absolute value) of the rate are shifted as
compared to the standard case \cite{krauss}. For larger values of
$\vmin$ the distortion is less pronounced, and it dies away when
$\vmin \gsim 230$ Km s$^{-1}$.  This may be explained by the fact
that, as $\vmin$ grows, the integral of Eq.(\ref{eq:ivmin}) becomes
less sensitive to the parameter $\eta_x$ since it gets increasingly
dominated by WIMPs with velocities along the $y$ axis, which is the
one along which the boost due to the galactic rotation is directed.
We notice that for values of $\vmin$ around ($100-200$) Km s$^{-1}$ a
distortion is present, but the amplitude of the modulation is
suppressed and therefore difficult to detect.

\begin{figure}[t]
\centering
\vspace{-30pt}\includegraphics[width=1.0\columnwidth]{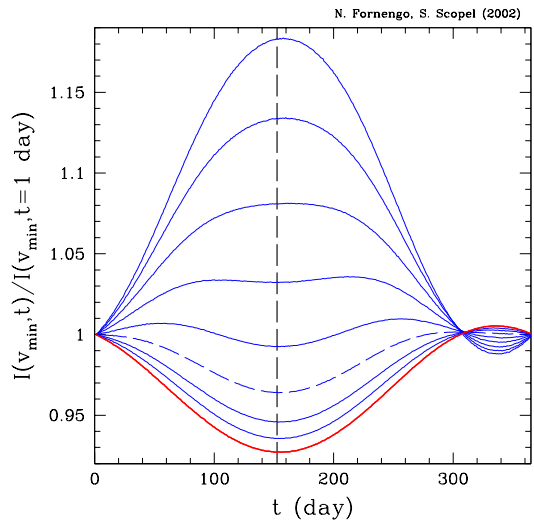}
\vspace{-30pt}\caption{\label{fig:ivmin_radial} The same as in
Fig. \ref{fig:ivmin_isotropic}, for an anisotropic model with
$\lambda_{12}\equiv\sigma_x/\sigma_y = 10$ and
$\lambda_{32}\equiv\sigma_z/\sigma_y = 3$. The curves refer to:
$\vmin=0$ (lower), 180, 190, 200 (dashed), 210, 220, 230, 240, 250
(upper) Km s$^{-1}$.}
\end{figure}

As a second example, in Fig.\ref{fig:ivmin_radial} we plot $\ivmin$ as
a function of time for the case $\lambda_{12}=10$, $\lambda_{32}=3$
(i.e.: $(\sigma_x,\sigma_y,\sigma_z) = (257,26,77)$ Km s$^{-1}$. This
situation is representative of a radial anisotropy.  In this case, we
have further enhanced the contribution of $\delta\eta_y$ over
$\delta\eta_{x,z}$, so that one should expect to draw the same
conclusions as in the isothermal sphere case, with the usual
sinusoidal time--dependence of the rate and a phase close to $t_y \sim
t_0$. This is indeed the case, except for a narrow interval of values
of $\vmin$ around $\vmin \simeq$ 210 Km s$^{-1}$. In this range,
$\ivmin$ develops two maxima because, for that particular choice of
$\vmin$, there is an exact cancellation in the first term of the
expansion in $\epsilon_y=\delta\eta_y/\eta^0_y$, so that the term
${\cal O}(\epsilon_y^2)$ proportional to $\cos^2[\omega (t-t_y)]$ sets
in. This particular cancellation of the first term in the Taylor
expansion of $\ivmin$ happens also for the isothermal--sphere model,
but in that case the size of the quadratic term is strongly suppressed
because $\epsilon_y$ is much smaller. In a WIMP direct detection
experiment this effect would show up in a very peculiar way: a halving
of the modulation period of the rate in a narrow range of recoil
energies. Note, however, that in order to have some realistic chance
to detect this effect, it should show up in one of the experimental
energy bins just above threshold, where the highest signal/background
ratio is usually attained.

\begin{figure}[t] \centering
\vspace{-30pt}\includegraphics[width=1.0\columnwidth]{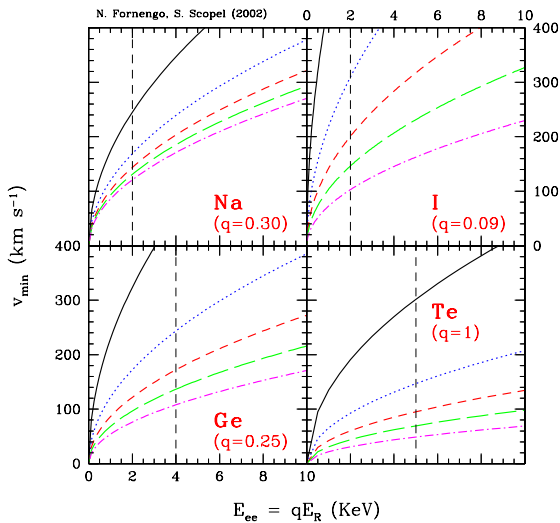}
\vspace{-30pt}\caption{\label{fig:vmin} Values of $\vmin$ as a
function of the quenched nuclear recoil energies $E_{ee}=q E_R$, for:
NaI scintillators \cite{damalast}, Ge ionization detectors
\cite{direct} and Te bolometers \cite{direct}. For each panel, the
different curves refer to WIMP masses of: 20 GeV (solid), 50 GeV
(dotted), 100 GeV (dashed), 200 GeV (long--dashed) and 1 TeV
(dot--dashed). The dashed vertical lines denote the current energy
thresholds.}
\end{figure}

In order to establish a link between our discussion and WIMP direct
detection experiments \cite{direct}, in Fig. \ref{fig:vmin} we plot
$\vmin$ as a function of the quenched nuclear recoil energy
$E_{ee}\equiv q E_R$ ($q$ is the quenching factor) for the target
nuclei: Na, I, Ge, Te and for different WIMP masses. The vertical
dashed lines show current energy thresholds achieved by each type of
detector. Fig. \ref{fig:vmin} shows that values of $\vmin \lsim$ 80 Km
s$^{-1}$, i.e.  sufficiently low to observe a sizeable distortion
effect as the one discussed for tangential anisotropy, correspond to
WIMP recoil energies below the threshold of present direct detection
experiments, and that the effect would be more easily detected at
higher WIMP masses. However, a foreseeable lowering of the threshold,
down to 0.5--1 KeV, could be enough to observe the distortion. On the
other hand, for radial anisotropy, we can conclude that the recoil
energy corresponding to a halving of the modulation period can
actually coincide to the experimental thresholds within the reach of
present--day detectors for 20 GeV $\lsim m_W\lsim$ 100 GeV, depending
on the particular target nucleus. In this respect, we note that the
properties of the annual modulation effect observed by the DAMA/NaI
experiment \cite{damalast} (a one--year--period sinusoidal behaviour
in the 2--6 KeV energy bins \cite{damalast,galaxy}) implies that the
DAMA/NaI experiment is already able to set constraint on strong radial
anisotropies.

\section{Conclusions}
In the present Letter we have shown that the main features of the
annual modulation of the signal of WIMP direct searches, i.e. the
sinusoidal dependence of the rate with time, the position of its
maxima and minima during the year and even the period, may be affected
by relaxing the isothermal sphere hypothesis in the description of the
WIMP velocity phase space. We have considered a multivariate gaussian
and found that different situations may occur, depending on the
pattern of anisotropy: tangential anisotropies induce a departure at
low energies from the usual sinusoidal time--dependence, along with a
shift in the position of the maximum of the signal during the year,
while radial anisotropies may produce a halving of the modulation
period in a particular energy bin. The former effect turns out to be
relevant at low recoil energies, actually below the threshold of
present--day experiments, while the latter should be already within
the reach of current detectors. In particular, the properties of the
annual modulation effect observed by the DAMA/NaI experiment
\cite{damalast} may already indicate that strong radial anisotropies
are excluded.

The effects discussed in this Letter may be used in the future to
provide a direct way of measuring (or setting limits on) the degree of
anisotropy of the galactic halo.


\end{document}